\documentclass[article, a4paper, pdftex]{siamart220329}

\usepackage{amsmath, a4, wasysym}  
\usepackage{graphicx}
\usepackage{hyperref}
\usepackage[utf8]{inputenc}
\usepackage{amssymb}

\usepackage[sort]{cite}

\usepackage{tabularx}
\usepackage{changepage}
\usepackage{booktabs}

\graphicspath{{figures/}}

\newcommand{\R}{\mathbb{R}}

\renewcommand{\phi}{\varphi}

\newcommand{\covid}{\textsc{COVID-19}}

\newcommand{\lpm}{\begin{pmatrix}}
\newcommand{\rpm}{\end{pmatrix}}

\newsiamremark{example}{Example}
\newsiamremark{remark}{Remark}
\title{Chaos in opinion-driven disease dynamics}
\author{Thomas Götz\thanks{Mathematical Institute, University Koblenz, Germany 
	(\email{goetz@uni-koblenz.de}; \email{pestow@uni-koblenz.de}; \email{moritzschaefer@uni-koblenz.de});).}
\and Tyll Krüger\thanks{Department of Information and Communication Technology, Faculty of Electronics, Wrocław University of Science and Technology, Wrocław, Poland
	(\email{tyll.krueger@pwr.edu.pl}).}
\and Karol Niedzielewski\thanks{Interdisciplinary Centre for Mathematical and Computational Modelling (ICM), University of Warsaw, Poland
		(\email{k.niedzielewski@icm.edu.pl}).}
\and Radomir Pestow\footnotemark[1]
\and Moritz Schäfer\footnotemark[1]
\and Jan Schneider\thanks{Faculty of Management, Wroclaw University of Science and Technology, Wrocław, Poland
		(\email{k.niedzielewski@icm.edu.pl})}}

\headers{Chaos in opinion-driven disease dynamics}{Thomas Götz, Tyll Krüger, Karol Niedzielewski, e.a.}

\begin{document}

\maketitle

\begin{abstract}
During the COVID-19 pandemic, it became evident that the effectiveness of applying intervention measures is significantly influenced by societal acceptance, which, in turn, is affected by the processes of opinion formation. This article explores one among the many possibilities of a coupled opinion-epidemic system. The findings reveal either intricate periodic patterns or chaotic dynamics, leading to substantial fluctuations in opinion distribution and, consequently, significant variations in the total number of infections over time. Interestingly, the model is exhibiting the protective pattern.
\end{abstract}


\section{Introduction}
Understanding the complex dynamics of opinion formation in a society is a challenging cross- disciplinary task where expertise and knowledge from many different fields is required. Mathematical models of opinion dynamics can provide quantitative and qualitative insights into the set of parameters and variables that shape and determine the dynamics and structure of opinions. Thanks to pioneers like Serge Galam in modelling social phenomena with methods borrowed and inspired by statistical physics whole new sub-fields of research like Sociophysics have been established ~\cite{galam_sociophysics_2012} and triggered subsequently work. Meanwhile there have been a large  variety of models designed like the Galam Models~\cite{galam_sociophysics_2008}, Voter model~\cite{castellano_nonlinear_2009}, Sznajd model~\cite{sznajd-weron_opinion_2000}, threshold models~\cite{granovetter_threshold_1978} and \textit{Bounded Confidence} models~\cite{hegselmann2002opinion} to name just a few.

Mathematical models for the epidemic dynamics of infectious diseases have a more than hundred years old tradition, starting with the work of Bernoulli in 1766 on smallpox inoculation~\cite{dietz_bernoulli_2000} and -- after a long pause -- Ross and Hudson on Malaria at the beginning of the last century ~\cite{noauthor_application_1917}.

But until the \covid{} pandemics, there was relative little interest among the epidemiological modeling community to incorporate aspects of opinion dynamics into epidemic models. Some notable exceptions are ~\cite{epstein_coupled_2008, wu_impact_2014, wu_dynamics_2016, pires_dynamics_2017, ni_modeling_2011, durham_deriving_2012, donofrio_information-related_2009, greenhalgh_awareness_2015}. During the  \covid{} pandemics it became evident, that the efficiency of the implementation of so-called Public Health and Social Measures (PHSMs) depend crucially on the level of acceptance in a society and  is therefore impacted by opinion formation processes. Furthermore, in most countries of the Western hemisphere, we have seen strong polarization dynamics within the society and deep digression on what is "the right thing to do" to fight the pandemic. These processes shaped also the attitude towards \covid{} vaccination and were causal for the large number of deaths in autumn 2021 due to insufficient vaccine uptake in the elderly population in eastern European countries like Poland and Bulgaria (for major works on coupled opinion-epidemic systems since 2008 see ~\cite{epstein_coupled_2008, sooknanan_trending_2020, sooknanan_fomo_2023, wu_dynamics_2016, pires_dynamics_2017, fang_coevolution_2023, wu_impact_2014, ali_impact_2023, zanella_kinetic_2023, ni_modeling_2011, agusto_impact_2023, kastalskiy_social_2021, bernardes_information_2021, sooknanan_harnessing_2021, jankowski_role_2022, du_how_2021, peng_multilayer_2021, durham_deriving_2012, carballosa_incorporating_2021, epstein_triple_2021, donofrio_information-related_2009, greenhalgh_awareness_2015, wagner_societal_2023} ).

In this article we study one of the many possibilities of a coupled opinion-epidemic system. The opinion dynamics we consider is a slight generalization of the so-called $q$--voter model (a kind of nonlinear voter process which favours the most prevalent opinion). The number of possible opinions is large respectively a continuum in our setting. For the epidemic dynamics we consider a simple $SIS$ system. Opinions have a direct impact on the likelihood to get infected (one possible interpretation could be to associate opinions with the frequency of mask wearing) and cause therefore heterogeneity in the relative  share of infections among the different opinion carriers. In the other direction an infection increases the likelihood to change the opinion (more precisely the likelihood of an infected individual to change its opinion is proportional to the the relative share of infected of his opinion among all infected). It is easy to show that the coupled system has no stationary solutions in the general case (section 2.5) despite the fact that in the decoupled system both -- the opinion and the epidemic dynamics -- converge to stable equilibria. 

It is a widespread belief that pure opinion formation systems are unlikely to show chaotic behaviour ~\cite{lim_social_2016} although there are examples of chaos even in simple deterministic opinion models~\cite{borghesi_chaotic_2006}. We observe for the system presented in this article either complex periodic patterns or chaotic dynamics with large fluctuations in the distribution of the opinions causing substantial variations over time in the total number of infected. 
It has been known that differences in the perceived risk can have dramatic impact on epidemic dynamics~\cite{epstein_coupled_2008}, and coupled opinion- epidemic systems 'can exhibit dynamics that do not occur when the two subsystems are isolated from one another'~\cite{sooknanan_trending_2020}. For instance, previous works presented emergence of periodic dynamics in opinion-epidemics models where behavior dynamics induced an instability of the  endemic equilibrium through a supercritical Hopf bifurcation -- an  effect leading to large oscillations~\cite{donofrio_information-related_2009, greenhalgh_awareness_2015}. Furthermore, recent study presented chaotic behavior in a coupled model induced by delayed response of behaviour to epidemic variables~\cite{wagner_societal_2023}. 
Our opinion dynamics and the coupling to the epidemic dynamics differs from the above mentioned studies and does not involve delay response nor periodic external triggers like seasonality to induce a non-stationary dynamics. 

%%%%%%%%%%%%%%%%%%%%%%%%%%%%%%%%%%%%%%%%%%
\section{Materials and Methods -- Model description}
The examined dynamical system consists of two coupled dynamical systems: an epidemiological system and an opinion formation system. 

\subsection{Epidemiological system}

In the following, we consider an infectious disease dynamics under the impact of opinions which affect the likelihood to get infected in a closed population of size $N$. Opinions are described by a one-dimensional continuous variable $x\in [0,1]$. $0$ value is reflecting point of view in opposition or polarization to the opinion with value $1$. Values between $0$ and $1$ reflect non-extreme opinions. Each individual is required to have one opinion $x$. Let $S(t,x)$ and $Z(t,x)$ denote the number of susceptible or infectious individuals with opinion $x$ at time $t$. A straight-forward extension of the classical Kermack-McKendrick $SIS$-model, see~\cite{kermack_contributions_1991}, leads to the system
\begin{subequations}
\label{eq:epi}
\begin{align}
    \frac {d} {dt} S(t,x) &= -\frac{\beta(x)}{N} S(t,x) \int_0^1 Z(t,y)\, dy + \gamma Z(t,x)\;,\\
    \frac {d} {dt} Z(t,x) &= \frac{\beta(x)}{N} S(t,x) \int_0^1 Z(t,y)\, dy - \gamma Z(t,x)\;.
\end{align}
Here, $\beta(x)>0$ denotes an opinion-dependent transmission rate. As a prototypical example, one may consider a linear dependence 
\begin{equation}
    \beta(x) = \beta^0 + (\beta^1-\beta^0) x\;,    
\end{equation}
\end{subequations}
where $\beta^0$ and $\beta^1$ denote the transmission rates for the two extreme opinions $x=0$ and $x=1$. However, it is advised that $\beta^0 \neq \beta^1$ to observe dynamics diverging from standard $SIS$-model. The recovery rate $\gamma>0$ is assumed to be independent of the opinion $x$.

\subsection{Opinion formation system}\label{sec:opiForm}
To describe the opinion dynamics, we employ a modification of the $q$-voter model~\cite{castellano_nonlinear_2009}. The classical $q$-voter model ($q>1$) for a population with just two mutually exclusive opinions reads as
\begin{equation}
    \label{eq:qvotersimple}
    v' = \alpha \left[ (1-v) v^q - v (1-v)^q \right] = \alpha v (1-v) \left[ v^{q-1}-(1-v)^{q-1}\right]\;,
\end{equation}
where $v$ denotes the fraction of individuals having one opinion and $1-v$ denotes the remaining individuals having the alternative opinion. A change of opinion occurs, if an individuals ''meets'' $q$ individuals of the other opinion. Then, with rate $\alpha$, the individuals flips to the other opinion. It is required that $\alpha>0$ since individuals are attracted to the opinion of others. It is obvious, that the simple $q$-model allows for the two opinion polarized equilibria $v=0$ and $v=1$ and the balanced equilibrium $v=\frac{1}{2}$. The two polarized equilibria are asymptotically stable, whereas the balanced equilibrium is unstable. To see this, we remark, that $v=0$, $v=1$ and $v=1/2$ are the roots of the right hand side $f(v)$ of the above ODE. Computing the derivative at these roots, we observe, that $f'(0) = f'(1)=-\alpha<0$ and $f'(1/2) = (q-1) \alpha (\tfrac{1}{2})^{q-1}>0$ showing the local instability or stability of the respective equilibria. 

Let $U(t,x)=S(t,x)+Z(t,x)$ denote the total number of individuals having opinion $x$, where $x\in [0,1]$ denotes a continuous range of opinions, e.g.~ranging between complete rejection of a non--pharmaceutical intervention ($x=0$) to full agreement with this intervention ($x=1$). Then $N=\int_0^1 U(t,x)\,dx$ equals to the overall population. In case of $q=2$ and a continuous spectrum of opinions $x\in [0, 1]$, a generalization of the classical $q$-voter model~\eqref{eq:qvotersimple} reads for $q=2$ as
\begin{multline}
\label{eq:opi}
    \frac{d}{dt} U(t,x) = (a U(t,x)^2 +\epsilon) \int_0^1 U(t,y)\, k(x,y)\, dy\\ - U(t,x) \int_0^1 (a U(t,y)^2+\epsilon)\, k(y,x)\, dy\;.
\end{multline}
Here, the non--negative kernel $k(x,y):[0,1]^2\to \R_+$ denotes the confidence (i.e. trust) of individuals of opinion $x$ and $y$ into each others judgment when interacted. With rate $a>0$ this interaction leads to a switch in the opinion. Typically, we will assume, that $k(x,y)=\rho(|x-y|) = \rho(r)$ depending just on the distance $r=|x-y|$ of the two opinions $x$ and $y$. It seems natural to assume, that $\rho$ is decaying with $r$, i.e.~the further apart the two opinions are, the less the trust into others judgment. The mechanism is called \textit{Bounded Confidence} (BC)~\cite{hegselmann2002opinion} in this work.

It is important to note that the probability of sampling of an individual with specific opinion in the continuous $x$ equals to zero. Therefore, in the system the interaction of individuals should be interpreted as the attraction of an individual to the opinion rather than "meeting" of other individuals.

The individual can either change its opinion according to $q$-individuals with rate $a$. The variable $\epsilon>0$ denotes a background rate of opinion change independent of encounters with other individuals. Proportion of actions are defined by parameters $a$ and $\epsilon$ with the assumption $0 \leq a + \epsilon \leq 1$. The proportion of individuals that do not take any action equals $1 - a - \epsilon$.

Some mathematical results for this model can be found in the appendix in propositions \ref{lem1} and \ref{lem2}.

\subsection{Opinion--epidemics Model}

In this section we will use $z(t,x)$ and\\ $u(t,x)$ as a densities of variables $Z(t,x)$ and $U(t,x)$ respectively (i.e. $z(t,x) = \frac{Z(t,x)}{\int_0^1 U(t,x)dx}$ and $u(t,x) = \frac{U(t,x)}{\int_0^1 U(t,x)dx}$). To couple the opinion dynamics with the infection process, we assume that the rate of changing one's opinion $x$ scales with the number of infected $z(t,x)$ having this opinion. The rational behind this assumption is the following: If an individual of opinion $x$ observes a large number of infected having the very same opinion $x$, the more likely it is, that the individual will change to another opinion $y$. And, in contrary, if there are only a few infected sharing ones opinion, then it is less likely to change the opinion. Scaling all populations with the total population $N$, incorporating the above idea into eqn.~\eqref{eq:opi} and combining it with the the $SIS$-model~\eqref{eq:epi}, we arrive at

\begin{subequations}
\label{eq:epi-opi}
\begin{align}
    \frac{d}{dt} z(t,x) &= \beta(x)\, \left( u(t,x)-z(t,x)\right) Z(t) - \gamma z(t,x)\;, \\
    Z(t) &= \int_0^1 z(t,y)\, dy\;, \\
    \beta(x) &= \beta^0 + (\beta^1-\beta^0) x\;, \\
    \frac{d}{dt} u(t,x) &= \frac{1}{Z(t)} \left( (a u(t,x)^2 +\epsilon) \int_0^1 z(t,y)\,u(t,y)\, \rho(|x-y|)\, dy \right.\\
        & \qquad \qquad\left. - u(t,x) z(t,x) \int_0^1 (a u(t,y)^2+\epsilon)\, \rho(|x-y|)\, dy \right)\;. 
\end{align}
Here, $Z(t)$ denotes the total number of infected.

A simple model for the interaction kernel $\rho(r)$ is the so--called \textit{Bounded Confidence} interval
\begin{equation}
    \rho(|x-y|) = \begin{cases}
        1 & \text{for } |x-y| \le \tau \\
        0 & \text{else}\;.
        \end{cases}\;,
\end{equation}
\end{subequations}
i.e.~if the difference between the two opinions $x$ and $y$ exceeds the \textit{Bounded Confidence} threshold $\tau$, no interaction occurs. 
Some analytic results for the asymptotic dynamics of the pure opinion dynamics are provided in appendix A and B . 

\subsection{Discretization of the Opinion Space}

To simulate the coupled\\ infection-opinion dynamics~\eqref{eq:epi-opi}, we discretize the opinion space $[0,1]$ by $n$ discrete opinions $x_1<\dots< x_n$. For simplicity, we assume to be equidistantly spaced $x_k = h(k-\frac{1}{2})$ for $k=1,\dots n$ and $h=1/n$. Let $z_i(t) = z(t,x_i)$ and $u_i(t)=u(t,x_i)$. Then, the opinion-discretized version of~\eqref{eq:epi-opi} reads as
\begin{subequations}
\label{eq:epi-opi-disc}
\begin{align}
    z_i' &= \beta_i\, \left( u_i-z_i\right) Z - \gamma z_i\;, \label{eq:zi}\\
    u_i' &= \frac{1}{Z} \left( (a u_i^2 +\epsilon) \sum_{k=1}^{n} z_k\,u_k\, \rho_h(|i-k|) - u_i z_i \sum_{k=1}^{n} (a u_k^2+\epsilon)\, \rho_h(|i-k|) \right) \;, \label{eq:ui}
\intertext{where}
    Z(t) &= \sum_{k=1}^{n} z_k(t)\;, \\
    \beta_i &= \beta^0 + (\beta^1-\beta^0) x_i\;,\\
    z_i' &= \frac d  {dt} z_i(t)\;,\\
    u_i' &= \frac d  {dt} u_i(t)\;
\end{align}
\end{subequations}
and $\rho_h(d)=\rho(hd)$ denotes the scaled \textit{Bounded Confidence} kernel.

The coupled $2n$-dimensional ODE system~\eqref{eq:zi}, \eqref{eq:ui} can be solved by any standard ODE solver, e.g.~a classical Runge-Kutta method with adaptive step sizes. For the simulation implementation details please see Section~\ref{sec:simulation}.

\subsection{Some analytic results for a simplified setting}

Consider the opinion-discretized system \eqref{eq:epi-opi-disc} in the simplified setting $\rho_h\equiv 1$ and $\epsilon=0$. We will show in the following that such a system cannot have a stationary solution except for the case when the likelihood to get infected does not depend on the opinion (in that case the system is actually decoupled).  Non-trivial equillibria for the infected compartments $z_i$ are characterized by the equation
\begin{equation*}
    z_i = \frac{\beta_i u_i Z}{\beta_i Z + \gamma}\,. 
\end{equation*}
Analogously, the equilibria for the opinion group $u_i$ have to satisfy
\begin{align*}
    u_i^2 \sum{z_k u_k} &= u_i z_i \sum_k u_k^2\;.
\intertext{Inserting the above relation for $z_i$  yields}
    \sum_k \frac{\beta_k u_k^2}{\beta_k Z + \gamma} &= \frac{\beta_i}{\beta_iZ+\gamma}\sum_k u_k^2\;.
\end{align*}
Hence, $\beta_i/(\beta_i Z+\gamma)$ has to be a constant, independent of $i$. This is only possible, if $\beta_i=\beta$ constant independent of $i$. 

Assuming $\beta_i=\beta$ to be independent of $i$, we define $A=\frac{\beta Z}{\beta Z+\gamma}$ and hence $z_i=Au_i$. Therefore,
\begin{align*}
    Z &= \sum_i z_i = \frac{\beta Z}{\beta Z +\gamma} \sum u_i
\intertext{and thanks to $\sum_i u_i =1$, we arrive at $1= \beta / (\beta Z +\gamma)$ or}
    Z &=1-\frac{\gamma}{\beta}\;.
\end{align*}
Feasible solutions $Z<1$ can only exist, if $\beta>\gamma$, i.e.~if the classical epidemiological threshold condition $\mathcal{R}_0=\frac{\beta}{\gamma}>1$ for the basic reproduction number $\mathcal{R}_0$ is satisfied.
The result is likely to hold for the general case ($\rho_h \neq 1$ and $\epsilon>0$) and epsilon not too large, but we where unable to obtain conclusive analytic results in this direction so far.

\subsection{Simulation setup}\label{sec:simulation}
\subsubsection{Software}
We run simulations using Julia language~\cite{bezanson_julia_2017} version 1.9.4. To resolve systems numerically we use \textit{DynamicalSystems}~\cite{datseris_dynamicalsystemsjl_2018, datseris_nonlinear_2022} and \textit{OrdinaryDiffEq}~\cite{rackauckas_differentialequationsjl_2017}. For the postprocessing and analysis \textit{ChaosTools}~\cite{datseris_nonlinear_2022}, \textit{StatsBase} and \textit{FFTW}~\cite{frigo_design_2005} packages are used.
\subsubsection{Hardware}
Computations are performed on \textsc{Cray XC40} (Okeanos) which is part of the ICM computing infrastructure. The system is composed of 1084 computing nodes. Each node has 24 \textsc{Intel Xeon E5-2690 v3} CPU cores with a 2-way Hyper Threading (HT) with 2.6 GHz clock frequency.

\subsubsection{Simulations}
Simulations were conducted in a discretized opinion\\ space $x$ with space points $n$. Simulations run for $30$,$000$ time steps with resolution 1. To allow for the system to stabilize its dynamics, initial $20$,$000$ steps are discarded and last $10$,$000$ steps are used for further investigation.

The coupled $2n$-dimensional ODE system~\eqref{eq:zi} is solved using Verner's “Most Efficient” Runge-Kutta method with order 6(5)~\cite{verner_numerically_2010}.

To analyze the system and find parameter sets with chaotic dynamics, we run simulations with parameter sampling in a grid search manner. We decided to alter values of parameters $n$, $\epsilon$, $\tau$ in equation \eqref{eq:epi-opi}. The parameters $a$, $\gamma$, $\beta_0$, $\beta_1$ remained constant. The setup allowed to vary and evaluate the influence of: the number of discretization space points $n$, the proportion of individuals that change opinion in unstructured manner $\epsilon$, and the \textit{Bounded Confidence} interval $\tau$ that controls the limits of mixing of the opinion in the population. We found these parameters to be the most important for the process. We set $a$ equal to $0.6$ to account for the fact that most people are conformist and follow the majority opinion. To make the system over-critical, we set limits of the function $\beta(x)$ (i.e., $\beta_0$ and $\beta_1$) to be above the $\gamma$ value. This way, the classical epidemiological threshold condition $\mathcal{R}_0=\frac{\beta}{\gamma}>1$ is satisfied. Values of $\beta_0$ and $\beta_1$ equal to $0.11$ and $0.225$, respectively, were chosen to ensure increase of transmission with increasing $x$. The initial opinion distribution was uniform $u(t=0,x) = 1.0$. The distribution of initially infected was uniform and equal to $0.01$ (i.e. $z(t=0,x)=0.01$).

Changeless parameters and their values of the equation are compiled in Table~\ref{tab:changeless}.
Changeable parameters and their range characteristics are compiled in Table~\ref{tab:changeable}.

\begin{table}[H]
\caption{Fixed parameters and their values used in a grid search.}\label{tab:changeless}
\newcolumntype{C}{>{\centering\arraybackslash}X}
\begin{tabularx}{\textwidth}{CCCCC}
\toprule
\textbf{Parameter}	& $a$       & $\gamma$  & $\beta_0$ & $\beta_1$ \\
\midrule
\textbf{Value}      & 0.6       & 0.1       & 0.11      & 0.225 \\
\bottomrule
\end{tabularx}
\end{table}

\begin{table}[H] 
\caption{Varied parameters and their ranges used in a grid search. \label{tab:changeable}}
\newcolumntype{C}{>{\centering\arraybackslash}X}
\begin{tabularx}{\textwidth}{CCCC}
\toprule
\textbf{Parameter}	        & $n$   & $\epsilon$	    & $\tau$    \\
\midrule
\textbf{Initial values}     & 4     & 0.0               & 0.15         \\
\textbf{Final value}        & 10 \textsuperscript{1}    & 0.4               & 1.05      \\
\textbf{Step size}          & 1     & 0.01		        & 0.1       \\
\bottomrule
\end{tabularx}
\noindent{\footnotesize{\textsuperscript{1} 20 is appended to the sequence in the end.}}
\end{table}%

\subsubsection{Analysis Methods}\label{sec:analysis}

Since we observed either periodic and chaotic behavior, we prepared a set of analysis methods allowing for an evaluation if system is periodic or chaotic and an insight on dynamics heterogeneity. We used the following methods:
\begin{itemize}
    \item autocorrelation
    \item maximum Lyapunov exponent (\textit{MLE})~\cite{lyapunov_general_1992,benettin_kolmogorov_1976}
    \item spectral Shannon entropy~\cite{haaga_juliadynamicscomplexitymeasuresjl_2023,llanos_power_2017,tian_spectral_2017,shannon_mathematical_1948}
    \item standard Shannon entropy~\cite{shannon_mathematical_1948}
    \item Poincaré maps
    \item Fourier Transform
\end{itemize}
Autocorrelation is a standard tool for detecting periodic signal and its frequency.
In the setup, we compute maximal value of autocorrelation of arbitrarly chosen lags ranging from 150 to 300. Autocorrelation value should be close to $1.0$ for the periodic dynamics.
Lyapunov exponents measure rates of separation of nearby trajectories in the flow of a dynamical system. In analysis we use \textit{MLE} to account for the maximal exponential separation. Positive \textit{MLE} indicates chaotic dynamics.
Spectral Shannon entropy is a measure of heterogeneity of signal that is computed on the basis of square of amplitudes of its Fourier transform. The values are than normalized to $1$ and standard Shannon entropy is computed. The closer the signal is to white noise, i.e. chaotic, the higher the entropy.
Standard Shannon entropy is used as a measure of heterogeneity of opinion dynamics. It is computed based on the densities of opinions in each time step. Then by obtaining maximum, minimum, mean and range length of entropies (i.e. difference of maximum and minimum) we get an insight into diversity of opinions.
Poincaré map is a standard descriptive method used to determine periodicity of the system. In our work we plot $u_i(t)$ vs $u_i(t+1)$ and $z_i(t)$ vs $z_i(t+1)$ to visualize orbits. When system is periodic it has evident closed loops. Fast Fourier Transform (FFT) is standard mathematical method of spectral analysis of discrete signals. It is a tool for describing time series with frequencies and their amplitudes.

%%%%%%%%%%%%%%%%%%%%%%%%%%%%%%%%%%%%%%%%%%
\section{Results and Discussion}

In this section we evaluate coupled model dynamics on account of its chaotic or periodic nature. We performed $2870$ simulations in a grid search manner to investigate the consistency, heterogeneity, variety and quality of the results. We present the numerical results along with their qualitative and quantitative analysis.

\begin{figure}[htb]
\centering
\includegraphics[width=0.9\textwidth]{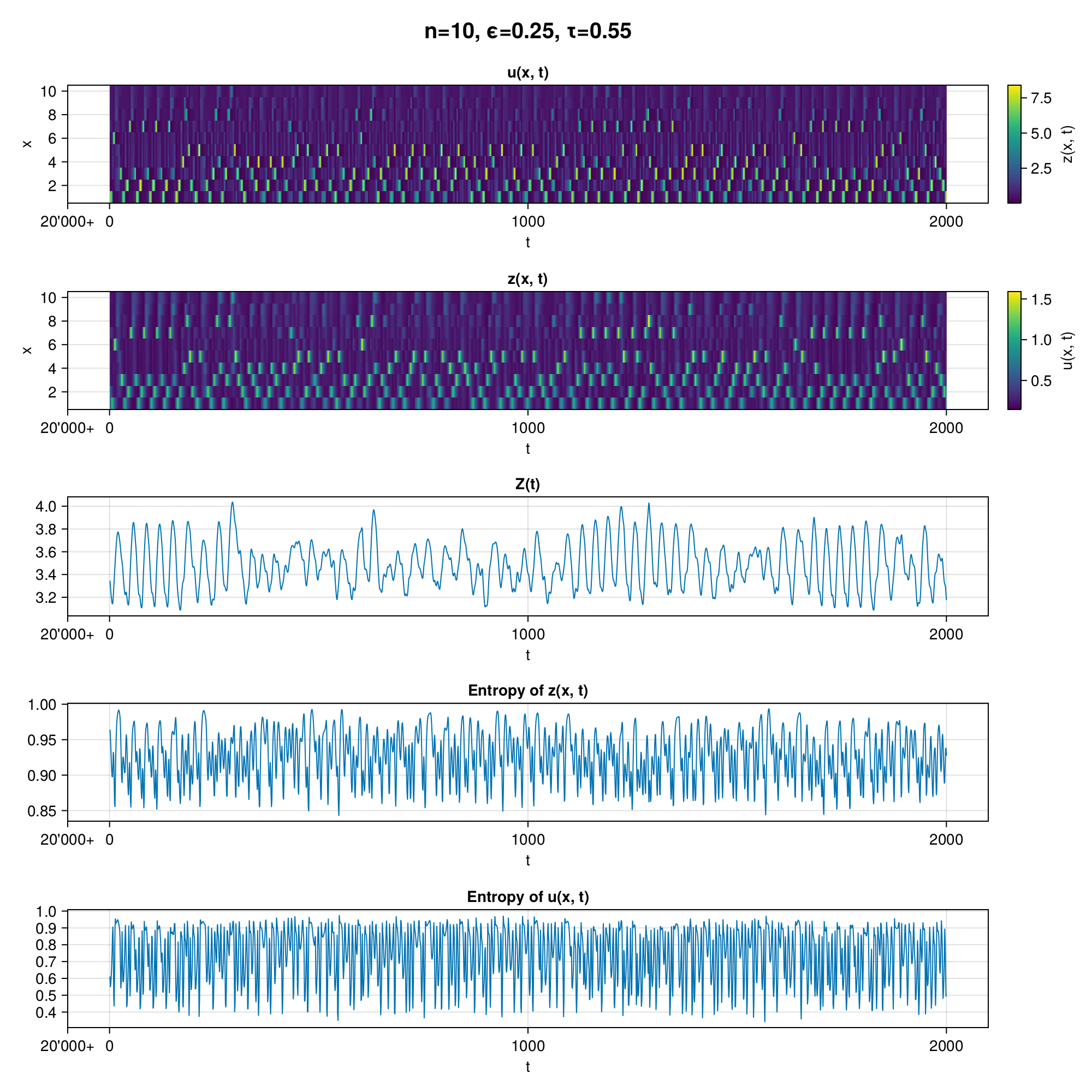}
\caption{Example of a chaotic timelines of the evolution of the dynamics with parameters $n=10,\epsilon=0.25,\tau=0.55$ where MLE and autocorrelation indicate chaotic behavior. The timelines begin with 20,000+ time steps and finishes with 20,000+2,000. We show heatmaps of opinion $u(x, t)$, infected $z(x, t)$, sum of infected $Z(t)$ and entropies of $z(x, t)$ and $u(x, t)$ repectively. Entropies of $z$ and $u$ are computed with with base 10. In each plot there are fluctuations and irregularity of data, especially in $u$ and entropy of $u$.}\label{fig:chaoticTimeline}
\end{figure}

\begin{figure}[htb]
\centering
\includegraphics[width=0.9\textwidth]{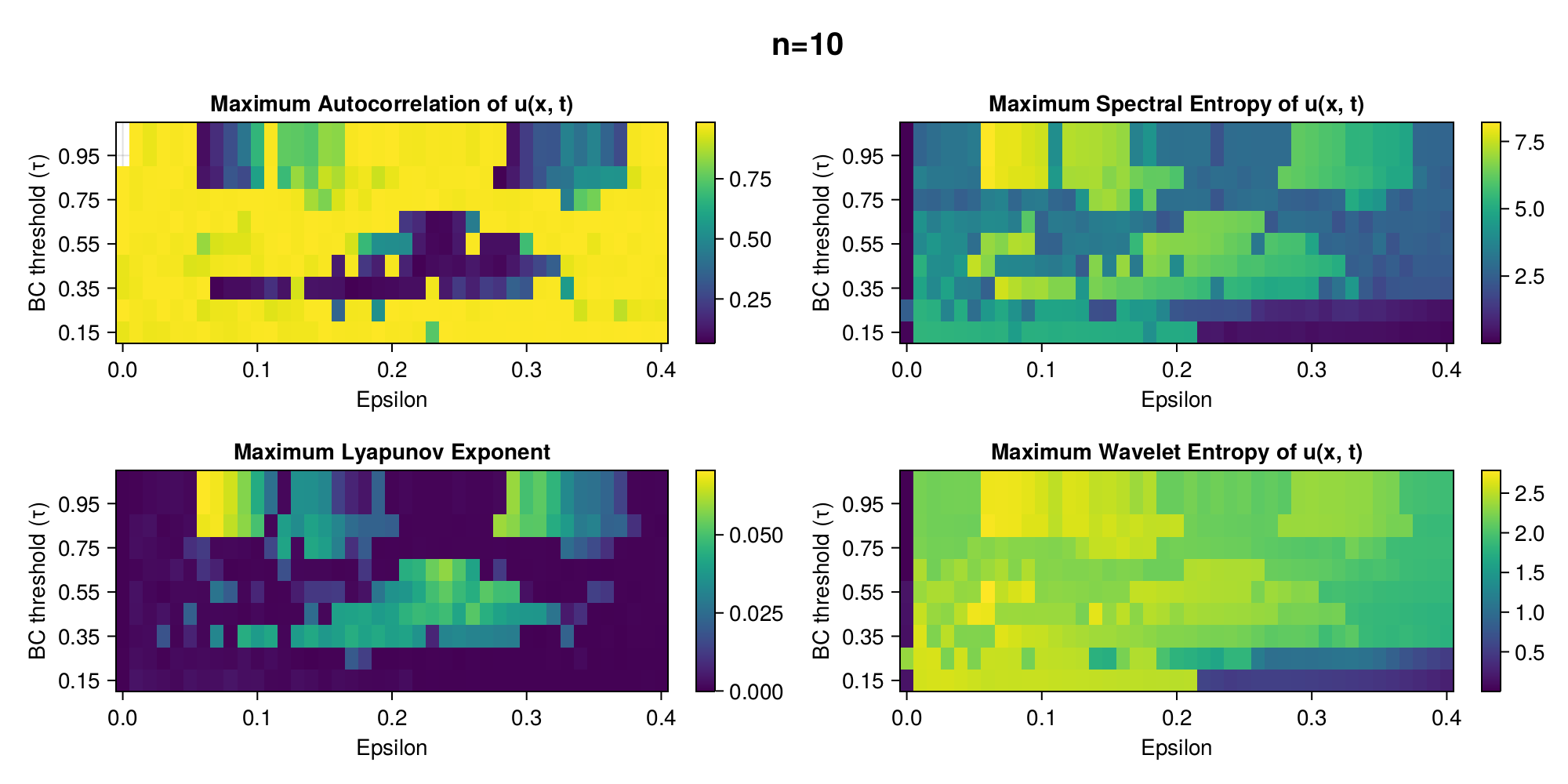}
\caption{Panel of heatmaps with measures of disorder of dynamics of system with $n=10$. The results are from the grid search simulations for $\epsilon$ and $\tau$ parameters that are on x and y axis respectively. Maximum Autocorrelation (top left), \textit{MLE} (bottom left), Maximum Spectral Entropy (top right), Maximum Wavelet Entropy (bottom right) are displayed. In each plot there is large fluctuation of values. Missing tiles in Autocorrelation heatmap are NaN values that mark stationary solutions. The parameter spaces with chaotic simulations are easy to distinguish in Autocorrelation (low values), \textit{MLE} (high values) and Spectral Entropy (high values). The low Autocorrelation values, high \textit{MLE}s values and high Spectral Entropy overlap. It is expected behavior that increases confidence into chaoticity in these regions. Maximum Wavelet Entropy is an outlier that is less consistent with other measures.\label{fig:n10}}
\end{figure}

In Figure~\ref{fig:chaoticTimeline} we can see visualization of simulation with parameters where \textit{MLE} and autocorrelation indicate chaotic regime. In the top two subplots we can see visualization of raw $u$ and $z$ data. The absence of clear recurrent patterns is a visual indicator for chaotic behaviour (in Figure~\ref{fig:n10} we see also that for the chosen parameter values we have positive Lyapunov exponents). It is apparent as well for the three bottom subplots that show aggregate $Z$, entropy of $z$ and entropy of $u$. We can see that $Z$ exhibit deviation of values within $\sim20\%$. Entropy of $u$ exhibit greater range of fluctuations than entropy of $z$. This might be due to the fact that $u$ does not translate to the values of $z$ as the non--uniform $\beta(x)$ function is used. Nevertheless, It is hard to notice any order in the timelines.

\begin{figure}[htb]
\centering
\includegraphics[width=0.9\textwidth]{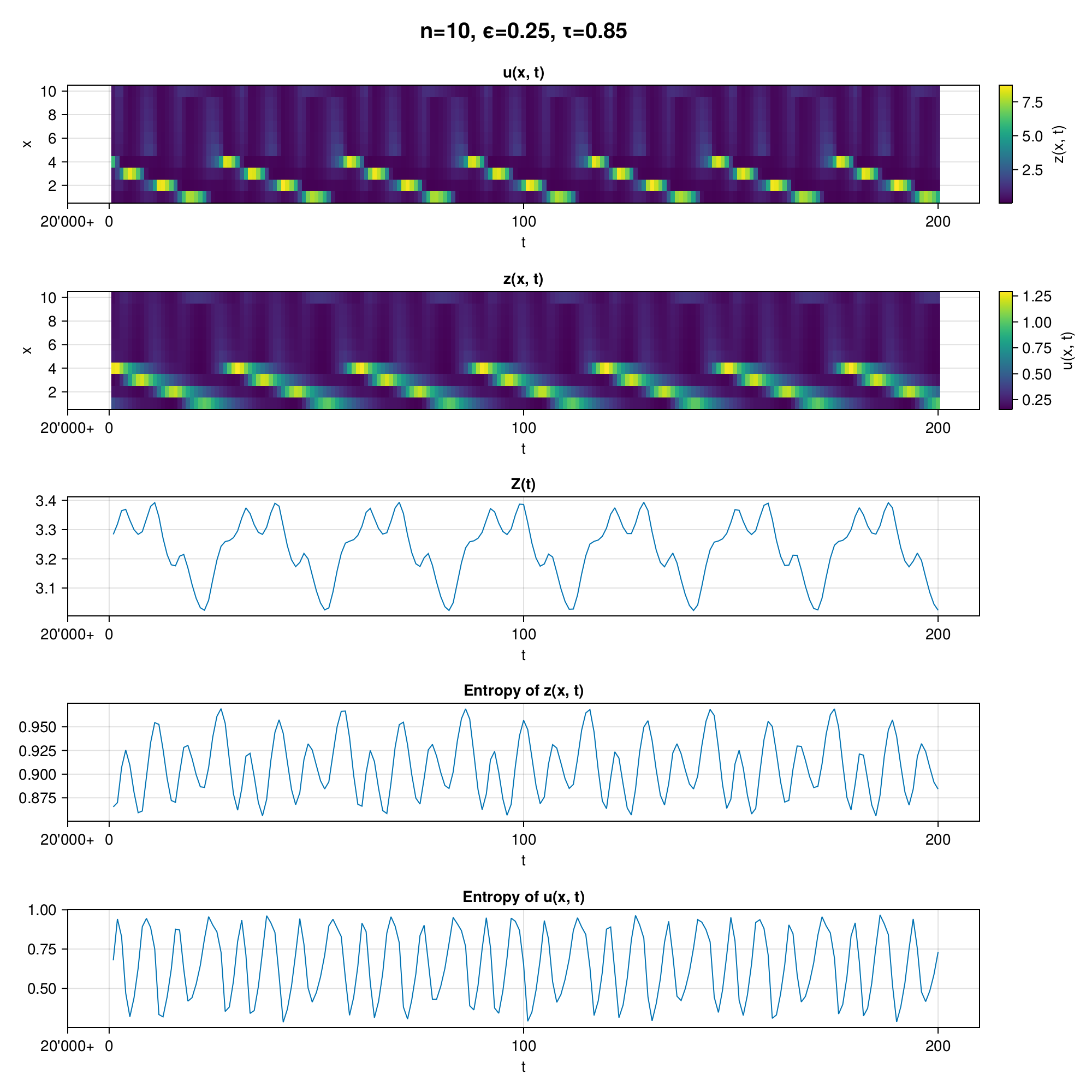}
\caption{Example of a periodic timelines of the evolution of the dynamics with parameters $n=10,\epsilon=0.25,\tau=0.85$ where \textit{MLE} and autocorrelation indicate periodic behavior. The timelines begin with 20,000+ time steps and finishes with 20,000+200. We show heatmaps of opinion $u(x, t)$, infected $z(x, t)$, sum of infected $Z(t)$ and entropies of $z(x, t)$ and $u(x, t)$ repectively. Entropies of $z$ and $u$ are computed with with base 10. In each plot there are fluctuations of data, especially in $u$ and entropy of $u$. One can clearly distinguish periodic behavior in each plot.\label{fig:periodicTimeline}}
\end{figure}

In Figure~\ref{fig:periodicTimeline} we can see visualization of simulation with parameters from periodic regime. In the top two subplots we can see visualization of raw $u$ and $z$ data. Recurrent pattern is easy to notice in this example. Namely, $u$ and $z$ are oscillating in a descending order. Even though the pattern seems simple, it exhibits complex dynamics observable in $Z$ and both entropies plots. Similarly to the first panel~\ref{fig:chaoticTimeline}, entropy of $u$ exhibit greater range of values than entropy of $z$. Contrary to previous panel, we can easily distinguish periodicity in each subplot.

\begin{figure}[htb]
\centering
\includegraphics[width=0.9\textwidth]{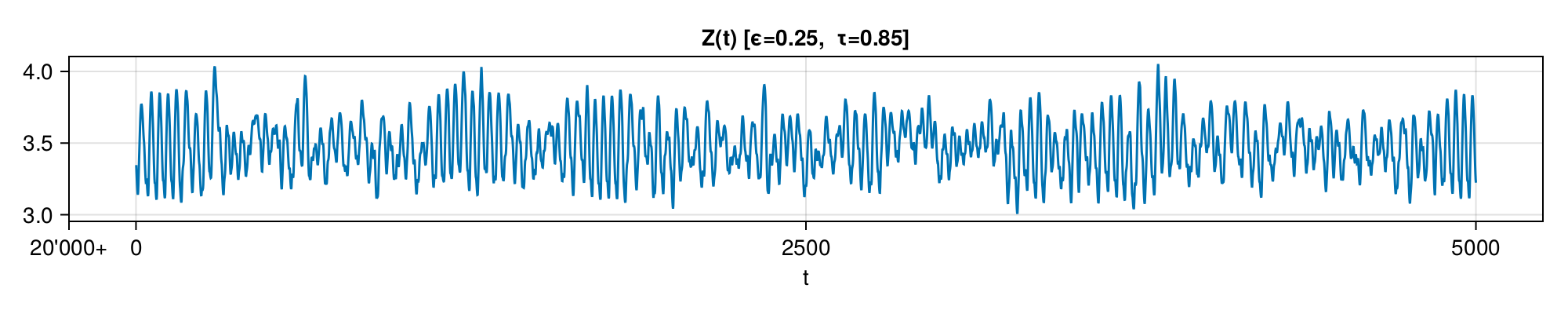}
\caption{Example of a chaotic timeline  of sum of infected $Z(t)$ from figure no.~\ref{fig:chaoticTimeline}. Simulation with parameters $n=10,\epsilon=0.25,\tau=0.55$ with extended time up to 20,000+5,000 steps is displayed.\label{fig:chaoticTimelineLong}}
\end{figure}

In Figure~\ref{fig:chaoticTimelineLong} we present extended disordered $Z$  dynamics from Figure~\ref{fig:chaoticTimeline}. Notwithstanding, it is evident that the dynamics of the system exhibit pseudo-periodic intervals that are intermittently disrupted by chaotic regimes. While the root cause of this chaotic behavior remains unexplored in this particular study, it serves as a potential hint for further investigations into the underlying pathways leading to chaos.

Till now, particular examples of numerical patterns of chaotic and periodic behavior were presented. Simulations have highly non--trivial chaotic behavior and non--trivial periodic behavior. To give a comprehensive perspective on the complexity of dynamics involved in the system, the assessment of disorder in simulations encompassing a diverse array of parameters are illustrated in subsequent figures.

\begin{figure}[htb]
\centering
\includegraphics[width=0.9\textwidth]{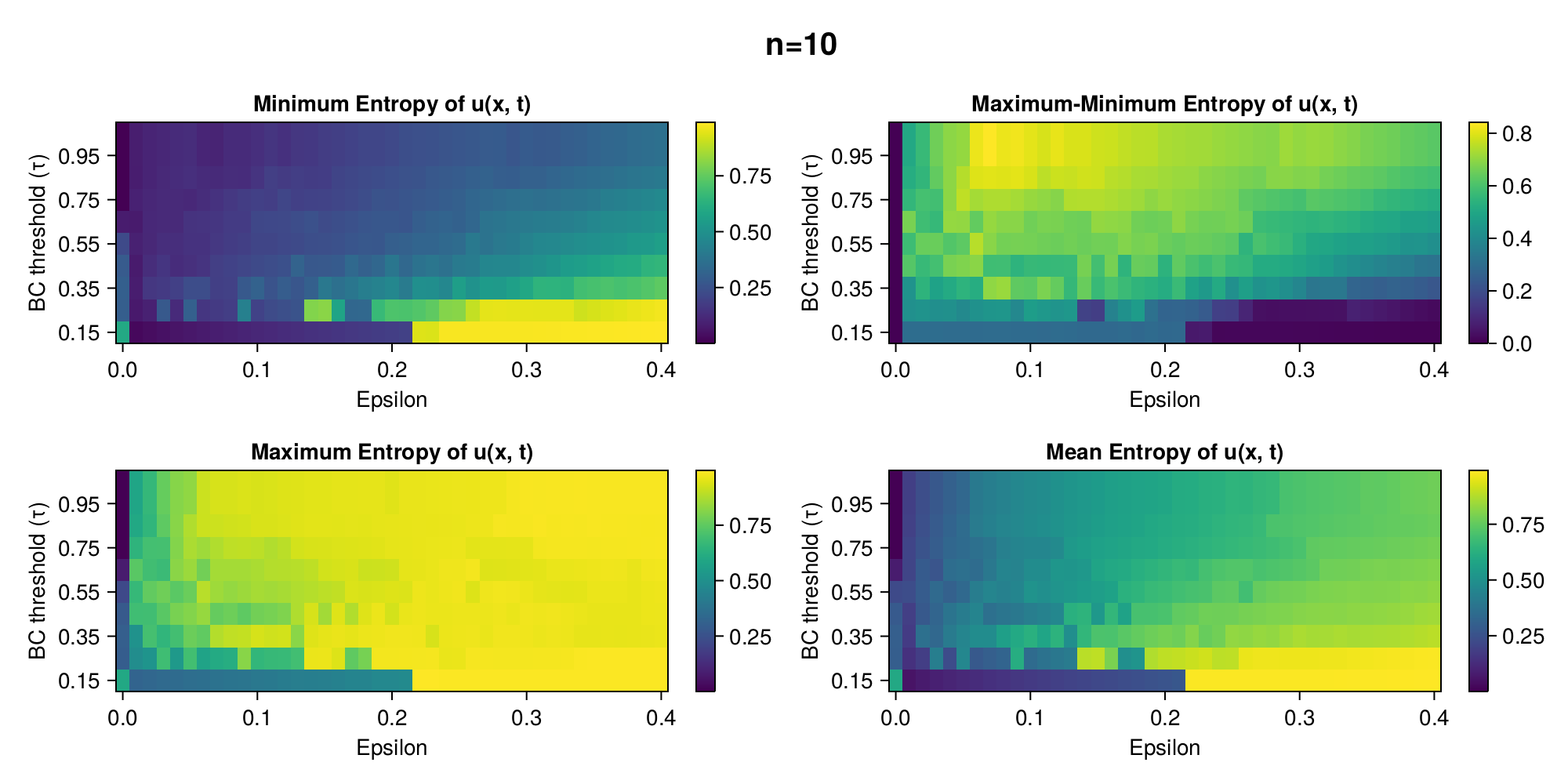}
\caption{Panel of heatmaps with measures of dispersion of dynamics of opinion $u(x, t)$ with $n=10$. The results are from the grid search simulations for $\epsilon$ and $\tau$ parameters that are on x and y axis respectively. Minimum Entropy (top left), Maximum Entropy (bottom left), difference of Maximum and Minimum Entropy (top right), Mean Entropy (bottom right) are displayed. In each plot there is large fluctuation of values. We can see the larger oscillations in opinion range the larger the BC threshold. For high epsilon and low BC we have very consistent and narrow opinion distribution. The highest oscillations in entropy match with the highest \textit{MLE} in figure~\ref{fig:n10}.\label{fig:n10UEntropy}}
\end{figure}

\begin{figure}[htb]
\centering
\includegraphics[width=0.9\textwidth]{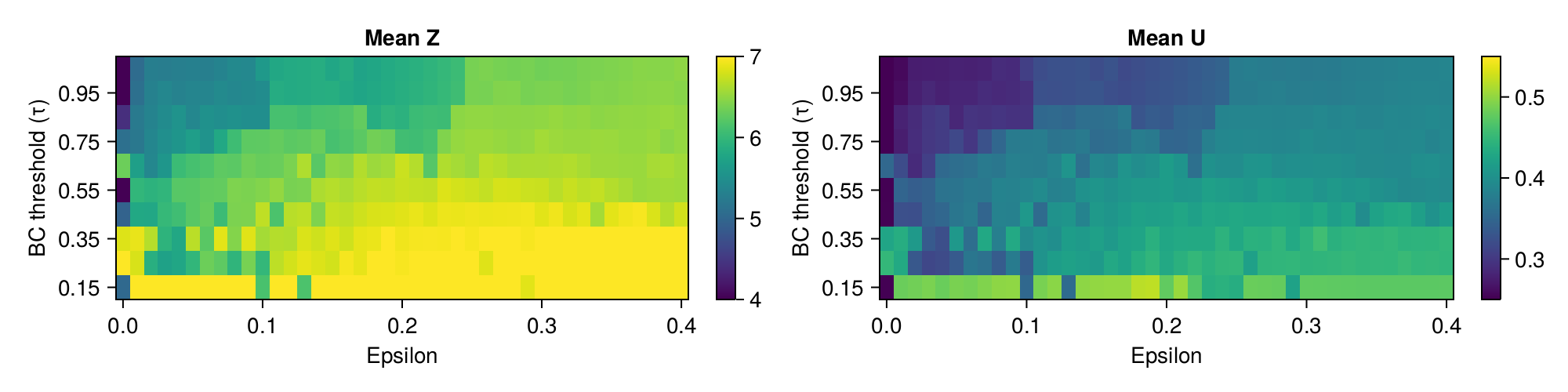}
\caption{Panel of heatmaps with measures of expected values sum of infected $Z(t)$ and of opinion $u(x, t)$ with $n=10$. The results are from the grid search simulations for $\epsilon$ and $\tau$ parameters that are on x and y axis respectively. Mean $Z$ (left), average opinion (right) are displayed. In each plot there is large fluctuation of values. The higher the $\tau$ the lower the mean of infected and mean opinion. The higher the $\epsilon$ the higher the mean number of infected and mean opinion. When Epsilon is largest and $\tau$ is smallest (bottom right corner) we can see the highest mean of infected. \label{fig:meanZU}}
\end{figure}

The heatmaps in Figures~\ref{fig:n10}--\ref{fig:meanZU} present variety of metrics described in Section~\ref{sec:analysis}.

The panel in Figure~\ref{fig:n10} consists of 4 heatmaps with values of Autocorellation, \textit{MLE}, Spectral Entropy and Wavelet Entropy. From figures no.~\ref{fig:chaoticTimeline} and~\ref{fig:periodicTimeline} we know that $u$ has greater entropy variation than $z$, therefore we use it to compute autocorrelation and entropies. Missing tiles in autocorrelation heatmap are NaN values that mark stationary solutions. The chaotic scenarios are easy to distinguish in Autocorrelation (low values), \textit{MLE} (high values) and Spectral Entropy (high values) heatmaps. We can see three non--uniform regions: two in the top and one in the centre. Overlap of all three is evident with some divergence on the perimeter of chaotic regimes. Convergence of three measures is a strong computational indicator of chaos in the system. Maximum Wavelet Entropy is an outlier displaying a level of consistency less aligned with the other measures. The chaoticity regions in this instance are subtly and delicately outlined.
To better visualize the fluctuations and correlation of Autocorrelation and \textit{MLE} we include cross section plots along x and y axes in the appendix~\ref{fig:crossSections}. 

In Figure~\ref{fig:n10UEntropy} there is a panel picture with 4 heatmaps with statistical properties of entropy of $u$. Namely: minimum entropy, maximum entropy, range length of entropy (difference of maximum and minimum value), mean entropy.
We can see the larger the BC threshold the larger oscillations in opinion range.
For high epsilon and low BC we have very consistent and narrow opinion distribution. The highest oscillations in entropy match with the highest \textit{MLE} in Figure~\ref{fig:n10}.

In Figure~\ref{fig:meanZU} there are 2 heatmaps with mean $Z$ and mean opinion $x$. The large BC threshold correlates with the low mean of infected and the low mean opinion. Inversely, the lower the $\epsilon$ the lower the mean of number of infected. This suggests a rational response to epidemics, wherein individuals prioritize self-protection, aligning their opinions with those in favor to a protective regime (lower values of opinion $x$). Notably when both mechanisms are involved with the strongest influence (largest $\epsilon$ smallest $\tau$) we can see the highest mean infected indicating the worst epidemic outcomes. Surprisingly, there is no visible influence of the chaoticity mode on the mean epidemics results. It remains uncertain whether the mode is irrelevant in general or the chaotic fluctuations are too subtle to have a noticeable impact on the epidemics.

\begin{figure}[htb]
\centering
\includegraphics[width=0.9\textwidth]{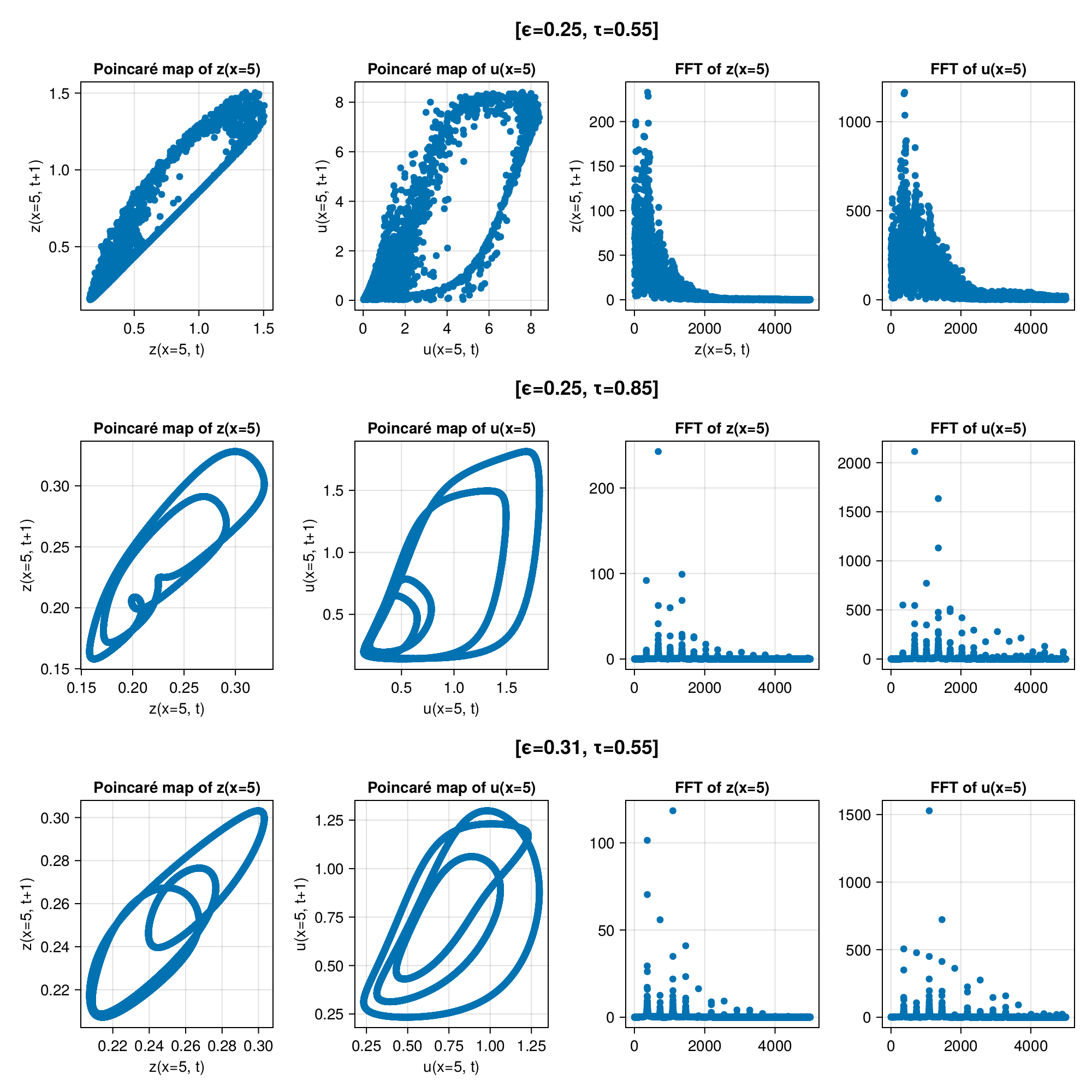}
\caption{Panel of descriptive statistical analysis plots of three simulations: 1 chaotic (top row) and 2 periodic (middle and bottom row). Top row present results from simulation with parameters $n=10,\epsilon=0.25,\tau=0.55$. Middle illustrates simulation with only $\tau$ parameter changed ($0.55\rightarrow0.85$). Bottom row of pictures illustrate simulation with only $\epsilon$1 parameter changed ($0.25\rightarrow0.31$). In first column there is Poincaré map of $z(x=5)$. In second column there is Poincaré map of $u(x=5)$. In third column there is Fast Fourier Transform of $z(x=5)$. In fourth column there is Fast Fourier Transform of $u(x=5)$. For sake of clear visualization we present only absolute amplitudes and half of the frequencies (to avoid symmetric picture). Poincaré maps have closed loops in middle and bottom row, contrary to the top one. Spectral analysis in top row is highly noisy in relation to the ones below it.\label{fig:poincare}}
\end{figure}

Panel no.~\ref{fig:poincare} depicts descriptive statistical analysis of three simulations: 1 chaotic and 2 periodic. Chaotic mode is present in the top row. The middle and bottom rows illustrate simulations with only $\tau$ parameter changed ($0.55 \rightarrow 0.85$) and with only $\epsilon$ parameter changed ($0.25 \rightarrow 0.31$) respectively. In the two columns on the left hand side we have Poincaré maps of $z(x=5)$ and $u(x=5)$. In the two columns on the right hand side there are Fast Fourier Transform results of $z(x=5)$ and $u(x=5)$. For sake of clear visualization we present only absolute amplitudes and half of the frequencies (to avoid symmetric picture). The difference is evident in panel plots. Poincaré maps have closed smooth loops in the middle and bottom rows, contrary to the top one. Spectral analysis in top row is highly noisy in relation to the ones below it. In the panel either periodic and chaotic simulations have highly complex dynamics. Periodic regimes demonstrate non--trivial dynamics even so there is lack of any noise. We can clearly see that using descriptive statistical methods we can easily distinguish the chaotic from periodic modes and detect evident noisiness of chaotic regime.

\begin{figure}[htb]
\centering
\includegraphics[width=0.9\textwidth]{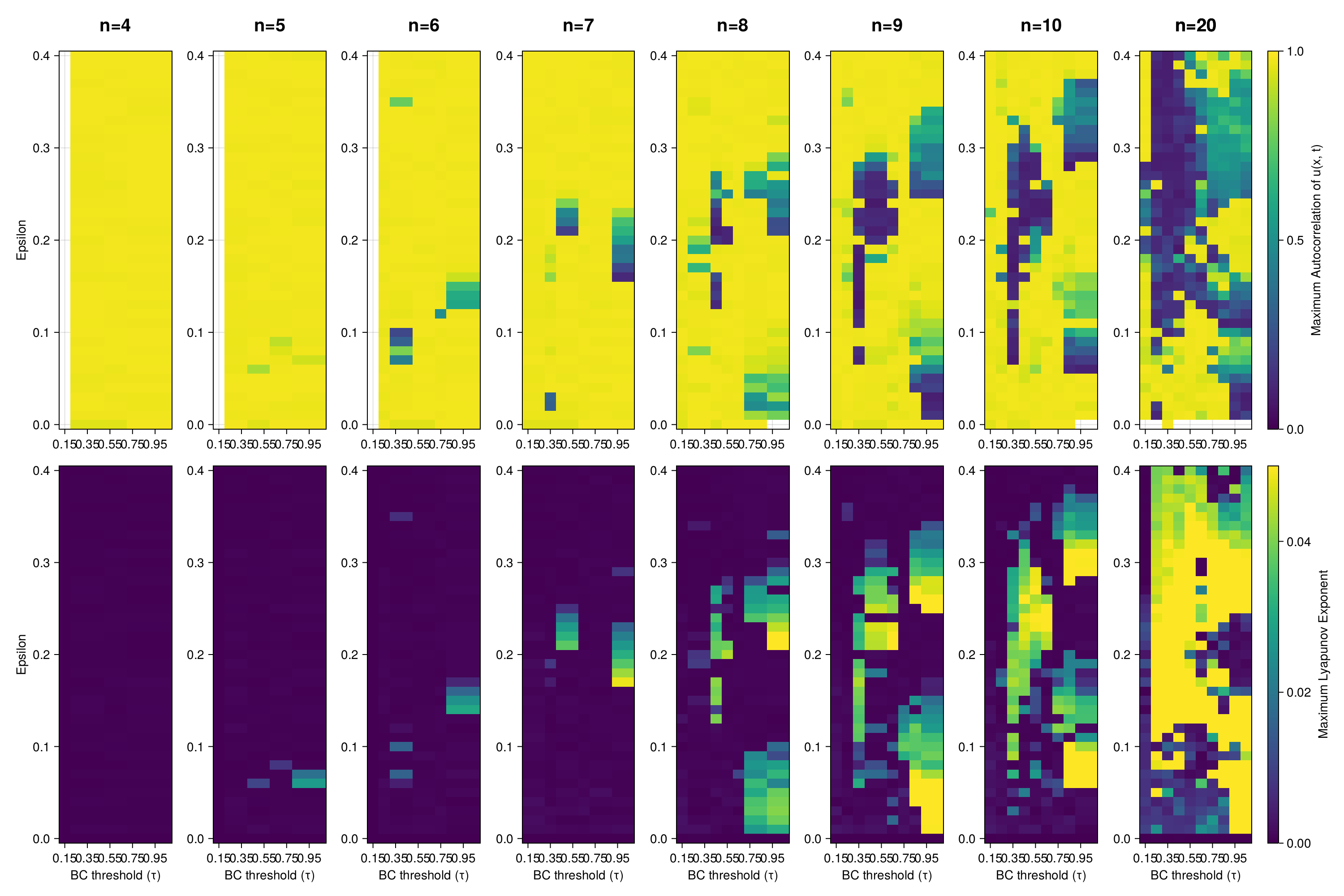}
\caption{Panel of heatmaps of Autocorrelation and \textit{MLE} measures according to various discretization resolution. First row consists of Autocorrelation heatmaps. Second row consists of \textit{MLE}. In columns we see results for resolutions ranging from $n=4$ to $n=20$. Missing tiles in Autocorrelation heatmaps are NaN values that mark stationary solutions. We can see that number of confined parameter spaces with chaotic regimes varies between $n$ values and the size of area of unordered dynamics occupied is increasing with $n$. The chaotic space (positive \textit{MLE}) takes up to $\sim90\%$ of total space for $n=20$.\label{fig:n4:20}}
\end{figure}

All presented simulation up to now had number of discretization points $n$ equal to $10$. In figure no.~\ref{fig:n4:20} we present heatmaps of autocorrelation and \textit{MLE} measures according to various discretization resolution ranging from 4 to 20 points. The minimal resolution for the chaotic regime to occur is 5. Missing tiles in autocorrelation heatmap are NaN values that mark stationary solutions. We can see that number of confined parameter spaces with chaotic regimes varies between $n$ values and the size of parameter space revealing chaotic dynamics is increasing with $n$. Most of the chaotic parameter regions have unstructured shapes and sizes. The pattern of forming of these areas is noisy and unclear. The positive \textit{MLE}s marking chaotic space takes up to $\sim90\%$ of total space for $n=20$.

\subsection*{Discussion}
It becomes evident that a minimum of 5 spatial points is required to observe chaotic behavior. The manifestation of chaotic dynamics is conditioned upon a sufficient number of states and \textit{Bounded Confidence} that facilitates the mixing of individuals with at least three different opinions. This outcome aligns with expectations, as when mixing only two states, there is just one possible opinion transition route (to the opposite one). As a result, opinion formation dynamics is strictly limited and periodicity is enforced.

Subsequently, when the \textit{Bounded Confidence} mechanism is deactivated (i.e., $\tau=1.05$), chaotic behavior cases are still observed. This suggests that while \textit{Bounded Confidence} may contribute to chaotic dynamics, it is not an obligatory factor for their occurrence.

On the contrary, all instances of chaotic dynamics emerge upon the presence of the $\epsilon > 0$. This underscores the essential role of opinion change independent of encounters as mandatory mechanism for the chaotic dynamics to develop.

An intriguing aspect of the model is the protective behavior embedded within it. We observe that the strong influence of independent opinion change and BC correlates with worse epidemics outcomes. Interestingly, in scenarios where communication across the population is widespread and uniform, and individuals follow the majority, the average number of infected cases is the lowest. 
Although we provided numerical evidence for chaotic dynamics only for systems with a finite number $n>4$ of opinions (up to $n=20$) we see no reason why chaotic patterns should become absent for large n respectively a continuum of opinions -- Figure~\ref{fig:n4:20} indicates that the opposite is true. A systematic numerical investigation of the continuous opinion system is challenging especially when search through  the parameter space is required (we observed runtimes of about 2 days for $n = 20$).

%%%%%%%%%%%%%%%%%%%%%%%%%%%%%%%%%%%%%%%%%%
\section{Conclusions and Outlook}
This study introduces a new coupled model for opinion and epidemic dynamics. Despite the absence of factors like seasonal effects, delayed response or contrarians which are known to trigger chaotic or complex dynamical pattern, our simulations and analyses reveal compelling evidence for both chaotic and complex periodic behavior. This observation is unexpected given that the constituent of the decoupled system is deterministic, exhibits only stationary dynamics, and the coupling mechanism is quite simple (infected individuals are just more likely to change their opinion than non-infected).

A closer examination of the chaotic timeline unveils pseudo-periodic intervals disrupted by a noisy signal. This phenomenon might be a hint for a possible route to chaos via intermittency. We did not found evidence for period doubling bifurcations. 

One implication of opinion--epidemics coupled systems chaotic behavior is the obvious difficulty to forecast dynamics beyond a time horizon larger than the inverse of the largest Lyapunov exponent~\cite{alfaro_forecast_2018}. 
There are several natural extension of our system. On the epidemic side it would be interesting to look at SIRS dynamics. Obviously many more ways of coupling the opinion and the epidemic dynamics are possible e.g. the opinion could not only impact the likelihood to get infected but also the infectivity of an individual. Finally there is a large number of opinion models studying the impact of heterogeneity of types of individuals in a society on opinion dynamics like the contrarians and conformists and it would be interesting to see how such systems coupled with epidemic dynamics behave. 

\acknowledgments{We would like to thank University of Koblenz and Wroclaw University of Science and Technology for providing the necessary infrastructure and scientific
environment for several meetings of the authors in Koblenz and Wroclaw within the DAAD–NAWA joint project.
Karol Niedzielewski and Jan Schneider would like to thank Infodemics Pandemics Summer School 2023 Lübeck for stimulating interest in coupled opinion and epidemics models.}

\conflictsofinterest{The authors declare no conflicts of interest.} 

\funding{This collaboration between the groups in Koblenz and in Poland was funded by the DAAD–NAWA joint project ”MultiScale Modelling and Simulation for Epidemics” – MSS4E, DAAD project number: 57602790, NAWA grant number: PPN/B-DE/2021/1/ 00019/U/DRAFT/00001}

\dataavailability{The code used in the research is available under \url{https://doi.org/10.5281/zenodo.10624921}.}

\abbreviations{Abbreviations}
{The following abbreviations are used in this manuscript:

\noindent 
\begin{tabular}{@{}ll}
BC  & Bounded Confidence\\
FFT &  Fast Fourier Transform\\
MLE & Maximum Lyapunov Exponent\\
MDPI & Multidisciplinary Digital Publishing Institute\\
DOAJ & Directory of open access journals\\
TLA & Three letter acronym\\
LD & Linear dichroism
\end{tabular}
}
\newpage
%%%%%%%%%%%%%%%%%%%%%%%%%%%%%%%%%%%%%%%%%%
%% Optional
% \appendixtitles{no} % Leave argument "no" if all appendix headings stay EMPTY (then no dot is printed after "Appendix A"). If the appendix sections contain a heading then change the argument to "yes".
\appendix

\section[\appendixname~\thesection]{}

\begin{Proposition}\label{lem1}
Consider the $q$-voter model 
\begin{align*}
\frac {d} {dt} u(t,x) &=u^q(t,x)\int u(t,y)\, dy - u(t,x) \int u^q(t,y)\, dy.
\end{align*}
Let $\tilde x=\text{argmax } u(t=0,x)$. Then $\tilde x=\text{argmax } u(t,x)$ for all times $t\geq 0$.
\end{Proposition}
\begin{proof}
Let the solution $u$ in the point $\tilde x$ be defined as $\tilde u(t):= u(t,\tilde x)$, and the solution in in an arbitrary point $x^*$ be defined as $u^*(t):=u(t,x^*)$. We assume $u^*(0)<\tilde u (0)$, so that $\Delta u(t):=\tilde u(t)-u^*(t)>0$ as well as $(u^*(t)+\Delta u(t))^q=\tilde u^q(t)$, resulting in the equation $u^{*q}(t)-\tilde u^q(t)=-\sum_{k=1}^q u^{*q-k}(t)\Delta u^k(t)$. It thus holds
\begin{align*}
\notag {u^*}'(t)-\tilde u'(t)&=(u^{*q}(t)-\tilde u^q(t))\int u(t,y)\, dy - (u^*(t) -\tilde u(t)) \int u^q(t,y)\, dy\\
&=-\sum_{k=1}^q u^{*q-k}(t)\Delta u^k(t) \int u(t,y) dy +\Delta u(t) \int u^q(t,y)dy
\end{align*}
It holds true that ${u^*}'(t)-\tilde u'(t)<0$ for $q=2$, since after division by $\Delta u(t)\neq 0$, it holds
\begin{align*}
\sum_{k=1}^2 u^{*2-k}(t)\Delta u^{k-1}(t)&=\sum_{k=1}^2 u^{*2-k}(t)(u^*(t)-\tilde u(t))^{k-1}\\
&=\tilde u(t) \int u(t,y) \,dy>\int u^2(t,y)\, dy.
\end{align*}
So it also holds ${u^*}'(t)<\tilde u'(t)$ for all $q\geq 2$, and thus $u^*(t)<\tilde u(t)$ for all $t \geq 0$. For $q\geq 2$, it can be shown by induction that it holds
\begin{align*} 
\sum_{k=1}^q u^{*q-k}(t)(u^*(t)-\tilde u(t))^{k-1}>\int u^q(t)\, dy,
\end{align*}
so that the statement holds for a general $q\geq2$-voter model as well.
\end{proof}

\begin{Proposition}[Stability of the Equilibrium in the pure $2$-voter model]\label{lem2}
 Regarding the system \eqref{eq:opi}, if $\epsilon$ is sufficiently large, i.e.~$\epsilon>a$, the uniform equilibrium $u(t,x)=1$ is stable.
\end{Proposition}
\begin{proof}
We consider a perturbation ansatz $u(t,x) = 1+ \delta e^{\lambda t} v(x)$, where $\int_0^1 v(x)\, dx=0$ for $\delta\ll 1$. If $\lambda<0$, then the uniform equilibrium $u\equiv 1$ is stable, otherwise not. Inserting the ansatz into system \eqref{eq:opi} we obtain
\begin{align*}
    \frac{d}{dt} u(t,x) &= a \left[ (u^2+\frac{\epsilon}{a}) \int_0^1 u\, dy - u\int_0^1 (u^2+\frac{\epsilon}{a})\, dy \right] \\
    &= a\left[ \left( 1+\frac{\epsilon}{a} + 2\delta e^{\lambda t} v(x)+ \delta^2 e^{2\lambda t} v^2(x)\right) \left( 1 + \delta e^{\lambda t} \int_0^1 v(y)\, dy\right) \right.\\
        &\hspace*{4em} \left.
        - \left( 1+\delta e^{\lambda t} v(x) \right) \int_0^1 1+\frac{\epsilon}{a} + 2\delta e^{\lambda t} v(y) + \delta^2 e^{2\lambda t} v^2(y)\, dy\right]\;.
\intertext{Linearizing with respect to $\delta$ and using $\int v\, dy=0$, we get}
    \frac{d}{dt} u(t,x) &= \delta\left(a- \epsilon\right) e^{\lambda t} v + \mathcal{O}(\delta^2)\;,
\intertext{and thanks to $\frac{d}{dt} u(t,x) = \delta \lambda e^{\lambda t} v$ we finally arrive at}
    \lambda &= \left( a- \epsilon\right) + \mathcal{O}(\delta)\;.
\end{align*}
For $\epsilon$ sufficiently large, i.e.~$\epsilon>a$, we get $\lambda<0$ and therefore the stability of the uniform equilibrium $u(t,x)=1$.

As one can see from the above derivation, this result also holds true for the generalization to a $q\geq2$-voter model.
\end{proof}

The above result shows the local stability of the equilibrium in the $2$--voter model. Proving global stability is a different issue in particular for the integro--differential model. The lack of knowlegde of a suitable Lyapunov function is a main obstacle when trying to analyze global stability. Therefore this issue is still subject of current research and beyond the scope of the given paper.

\section[\appendixname~\thesection]{}

\begin{figure}[htb]
\centering
\includegraphics[width=0.9\textwidth]{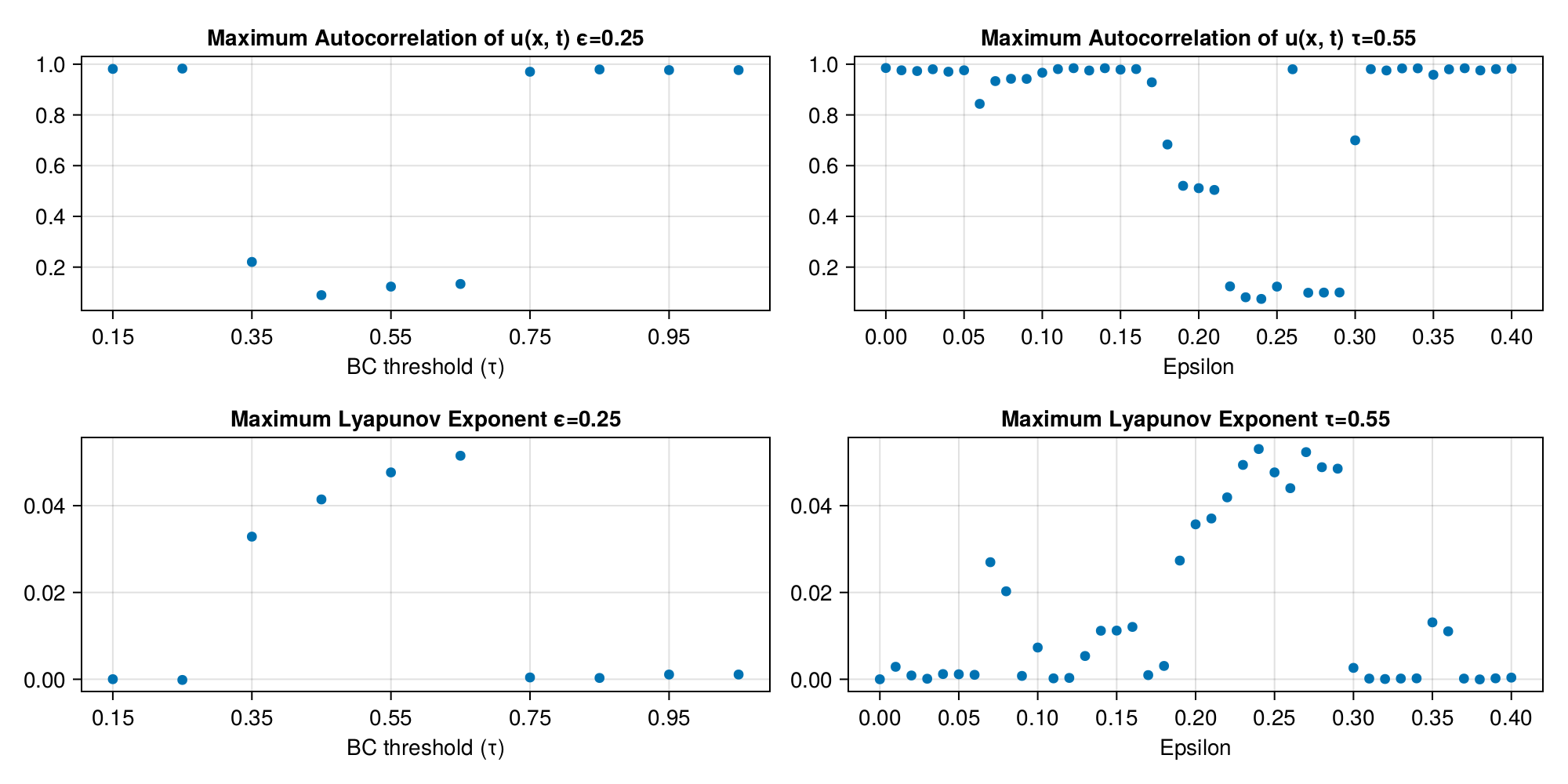}
\caption{Cross sections along Epsilon and $\tau$ axes of Autocorrelation and MLE heatmaps in figure no.~\ref{fig:n10}. Panel of 4 scatter plots includes: Maximum Autocorrelation along $\tau$ axis (top left), MLE along $\tau$ axis (bottom left), Maximum Autocorrelation along $\epsilon$ axis (top right),  Maximum Autocorrelation along $\epsilon$ axis (bottom right). We can see large fluctuations in each of 4 plots. The low Autocorrelation values correlate with high MLEs values when compared between top and bottom row. It is expected behavior that increases confidence into chaoticity in these regions. \label{fig:crossSections}}
\end{figure}


% \reftitle{References}


% \bibliography{bib.bib}
% \bibliographystyle{siam}

%=====================================
% References, variant B: internal bibliography
%=====================================

% \begin{thebibliography}{999}

\end{document}